\documentclass[aps,eqsecnum,twocolumn]{revtex4}
\usepackage{epsfig}

\newcommand{\eeq}{\end{equation}}
\newcommand{\ee}{\end{equation}}
\newcommand{\beq}{\begin{equation}}
\newcommand{\be}{\begin{equation}}
\def\bftheta{\mbox{\boldmath $\theta$}}

\begin{document}
%\draft

\title{Robust statistics for deterministic and stochastic\\
gravitational waves in non-Gaussian noise.  II: Bayesian analyses.} 

\author{Bruce Allen and Jolien D. E. Creighton}
\affiliation{Department of Physics, University of Wisconsin - Milwaukee,
P.O. Box 413, Milwaukee WI 53201}

\author{\'Eanna \'E.\ Flanagan}
\affiliation{Newman Laboratory of Nuclear Studies, Cornell University,
Ithaca, NY 14853-5001}

\author{Joseph D. Romano}
\affiliation{Department of Physical Sciences, University of
Texas at Brownsville, Brownsville TX 78520}

\begin{abstract}\quad
In a previous paper (paper I), we derived a set of near-optimal signal
detection techniques for gravitational wave detectors whose noise
probability distributions contain non-Gaussian tails.  The methods 
modify standard methods by truncating or clipping sample values which
lie in those non-Gaussian tails.  The methods were
derived, in the frequentist framework, by minimizing false alarm
probabilities at fixed false detection probability in the limit of
weak signals.  For stochastic signals, the resulting statistic
consisted of a sum of an auto-correlation term and a
cross-correlation term; it was necessary to discard ``by hand'' the
auto-correlation term in order to arrive at the correct, generalized
cross-correlation statistic.

In the present paper, we present an alternative
derivation of the same signal detection techniques from within the
Bayesian framework.  We compute, for both deterministic and stochastic
signals, the probability that a signal is present in the data, in the
limit where the signal-to-noise ratio squared per frequency bin is small, 
where the signal is nevertheless strong enough to be
detected (integrated signal-to-noise ratio large compared to one), and
where the total probability in the non-Gaussian tail part of the noise
distribution is small.  We show that, for each model considered, the
resulting probability is to a good approximation a monotonic function of the
detection statistic derived in paper I.  Moreover, for stochastic
signals, the new Bayesian derivation automatically eliminates the
problematic auto-correlation term.

\end{abstract}

\maketitle

%\newpage
%\narrowtext
%\twocolumn

\section{INTRODUCTION AND SUMMARY}
\label{s:intro}

Most of the literature on gravitational-wave data analysis assumes that the
detector noise is Gaussian.  However, significant non-Gaussian tails
have been a characteristic feature of the noise distributions in all
gravitational wave detectors constructed to date.  Standard detection
strategies for both deterministic and stochastic signals, which were
designed under the assumption of Gaussian noise, perform more poorly
when non-Gaussian noise is present.

In a previous paper in this series \cite{paperI} (henceforth paper I), 
we developed a new set of statistical signal-processing techniques to
search for deterministic and stochastic gravitational waves in
detector data.  These techniques are \emph{robust}, meaning that they
will work well even if the detector noise is not Gaussian but falls
into a broader statistical class that we expect includes realistic
detectors.  These new methods are similar to the older ones: one
constructs matched filters to search for known waveforms or
cross-correlates the instrument outputs at the different detector 
sites to search for a stochastic background.  The essential difference is that
the statistics are modified by truncation: detector samples
that fall outside the central Gaussian-like part of the sample distribution
are excluded from (or saturated when constructing) the
measurement statistic.  For both deterministic and stochastic signals,
a robust statistic was found which 
performs better than the optimal linear filter in the case where the detector
noise is non-Gaussian, and almost as well in the Gaussian-noise case.
Alternative methods for dealing with non-Gaussian noise for stochastic
signals have been explored by Klimenko and Mitselmakher \cite{KM}.

In paper I, we derived the statistics using the frequentist criterion
of minimizing false alarm probabilities at fixed false detection
probabilities in the limit of weak signals.  In the present paper, we
present an alternative derivation of the same signal detection
techniques from within the Bayesian framework.

We start in Sec.\ \ref{s:foundations} by reviewing the foundations of
the two different approaches to determining detection statistics used in
paper I and this paper.  We review the locally
optimal criterion used in paper I in Sec.\ \ref{ss:frequentist}.  In
Sec.\ \ref{ss:bayesian} we explain how Bayesian considerations lead to
a unique choice of detection statistic, as discussed by Finn
\cite{finn92}.  
In Secs.\ \ref{s:deterministic} and \ref{s:stochastic} of the paper,
we compute, for a variety of 
different models and sets of assumptions, that unique Bayesian
statistic.  We show that for each case considered, the Bayesian
statistic is equivalent to the statistics derived in paper I, in the
sense that the false alarm versus false dismissal curves of the two
statistics coincide to a good approximation.

The equivalence between the two types of statistic is valid
only under certain approximations, discussed below.  Under those
approximations, the Bayesian statistics which we obtain are equivalent to a
particular type of maximum likelihood statistic described in Refs.\
\cite{ML,steve}.  That type of maximum likelihood statistic differs
from the type of maximum likelihood statistic considered previously in
gravitational wave data analysis in its
treatment of noise  parameters.

Section \ref{s:deterministic} deals with known, deterministic
signals.   In Sec.\ \ref{ss:sdwnkw} we consider the case
of a known signal of 
unknown amplitude, incident on a single detector with white, Gaussian
noise, where the noise variance is assumed to be known.  For that case
the Bayesian statistic is shown to be equivalent to the standard
matched filtering statistic.  Section \ref{ss:sdwnkwu} generalizes
this analysis by allowing the noise variance to be an unknown
parameter, to be measured from the data; the same result is obtained.
In Secs.\ \ref{ss:color} and \ref{ss:sdcnkw} we consider the cases of
white and colored non-Gaussian detector noise.  For both of these
cases we show that the Bayesian statistic is equivalent to the
corresponding locally optimal statistic derived in paper I.

In Sec.\ \ref{s:stochastic} we give a similar analysis of stochastic
signals.  In Sec.\ \ref{ss:stochas1} we compute the Bayesian statistic 
for the case of a white stochastic signal,
and of two co-aligned detectors
with white Gaussian noise, where the noise variance is assumed to be
known.  The maximum likelihood statistic for this case was previously
computed, in a more general context, by Finn and Romano
\cite{FinnRomano}.   In this case we recover the result of Finn and
Romano: the optimal statistic is not the standard cross-correlation
statistic, but instead is a sum of the cross-correlation statistic and
and extra auto-correlation terms.  In Sec.\ \ref{ss:stochas2} we show
that, under the more realistic assumption where the noise variances
are taken to be unknowns to be 
determined from the data, then the standard cross-correlation
statistic is recovered.  Finally, in Sec.\ \ref{ss:stochas3} we
consider two detectors with white, non-Gaussian nosie, and we re-derive
the generalized cross-correlation statistic of paper I.
Section~\ref{s:conclude} contains a short conclusion and summary.

\section{FOUNDATIONS}
\label{s:foundations}

We denote the output of a set of gravitational wave detectors by
a vector ${\bf x}$, with
\be
{\bf x} = {\bf n} + {\bf s}.
\label{additivenoise}
\eeq
Here ${\bf n}$ is the detector noise, and ${\bf s}$ is a possibly present
gravitational wave signal.  We can write 
\be
{\bf s} = \epsilon {\hat {\bf s}},
\label{epsilondef}
\ee
where $\epsilon$ is a parameter governing the
signal strength, and the magnitude of ${\hat {\bf s}}$ is fixed.  As in
paper I, we shall be specializing to weak signals and using an
expansion in powers of $\epsilon$ about $\epsilon=0$ throughout this
paper.    

\subsection{Frequentist signal detection}
\label{ss:frequentist}

A key quantity is the probability distribution for ${\bf x}$
given $\epsilon$, $p({\bf x} | \epsilon)$.  This quantity can be used
to compute the performance of detection statistics in the frequentist
framework.  Suppose one is given a statistic
\be
\Gamma = \Gamma({\bf x}),
\eeq
say, and that one's detection criterion is that a signal is present if
$\Gamma({\bf x}) > \Gamma_*$ and not present otherwise, where
$\Gamma_*$ is a threshold.  Then the false dismissal probability
associated with this statistic is 
\be
\alpha(\Gamma_*) = \int_{\Gamma >\Gamma_*} d {\bf x} \, p({\bf x} | 0),
\label{alphadef}
\eeq
and the false alarm probability is
\be
\beta(\Gamma_*,\epsilon) = \int_{\Gamma <\Gamma_*} d {\bf x} \, p({\bf
x} | \epsilon).
\label{betadef}
\eeq
By eliminating $\Gamma_*$ between Eqs.\ (\ref{alphadef}) and
(\ref{betadef}) we obtain the false-dismissal versus false-alarm curve
\be
\beta = \beta(\alpha,\epsilon)
\eeq
which characterizes the performance of the statistic.

In paper I we showed that there is a unique statistic
$\Lambda_{(1)}({\bf x})$ which minimizes
$d \beta / d \epsilon$ at $\epsilon=0$ for fixed $\alpha$, defined by
the expansion  
\be
p({\bf x}|\epsilon) = p({\bf x}|0) \left[ 1 + \epsilon
\Lambda_{(1)}({\bf x}) + \epsilon^2 \Lambda_{(2)}({\bf x}) +
O(\epsilon^3) \right].
\eeq
This statistic therefore has the best 
false-dismissal versus false-alarm curve
for weak signals.  [If $\Lambda_{(1)}({\bf x})$ vanishes identically,
then $\Lambda_{(2)}({\bf x})$ is the unique statistic that minimizes 
$d^2 \beta / d \epsilon^2$ at $\epsilon=0$ for fixed $\alpha$.]  We
applied this class of detection statistics (called locally-optimal
statistics \cite{kassam}) to a variety of different gravitational-wave
signal detection problems.

\subsection{Bayesian signal detection}
\label{ss:bayesian}

In the Bayesian framework, the probability $P^{(1)}$ that a signal is
present in the data is given
\be
{P^{(1)} \over 1 - P^{(1)}} = \Lambda({\bf x}) {P^{(0)} \over 1 - P^{(0)}},
\label{fundamental}
\eeq
where $P^{(0)}$ is the {\it a priori} probability that a signal is
present, and $\Lambda({\bf x})$ is the likelihood ratio. 
In the literature on Bayesian statistics $\Lambda({\bf x})$ is called
a {\it Bayes factor}.  The Bayesian framework uniquely determines a
detection criterion, 
which is to threshold on the probability $P^{(1)}$ that a signal is
present.  From Eq. (\ref{fundamental}) it is clear that $P^{(1)}$ is a
monotonic function of $\Lambda({\bf x})$, so one can equivalently
threshold on $\Lambda({\bf x})$.  Thus, an optimal detection statistic
is uniquely determined in the Bayesian framework; however, that statistic does
depend on choices of prior probability distributions.

We now describe how likelihood ratio is computed.  It is given by the
formula 
\be
\Lambda({\bf x}) = \int d\epsilon \, \Lambda({\bf x},\epsilon)
p^{(0)}(\epsilon)
\label{Lambdaformula0}
\eeq
where $p^{(0)}(\epsilon)$ is the prior probability distribution for the
signal strength $\epsilon$, and
\be
\Lambda({\bf x},\epsilon) = {p({\bf x}| \epsilon) \over p({\bf x} | 0)}.
\label{Lambdaformula}
\ee
Suppose that the noise ${\bf n}$ is described by a probability
distribution $p_{\bf n}({\bf n})$, and the signal ${\bf s}$ by a
signal distribution $p_{\bf s}({\bf s} |  \epsilon)$.  Then it
follows from Eq.\ (\ref{additivenoise}) that 
\beq
p({\bf x} | \epsilon) = \int d {\bf s} \, p_{\bf n}({\bf x} - {\bf s})
\, p_{\bf s}({\bf s} | \epsilon).
\label{finn}
\eeq
The formula
(\ref{Lambdaformula}) can therefore be written as
\beq
\Lambda({\bf x} , \epsilon) = \int d {\bf s} \, { p_{\bf n}({\bf x} -
{\bf s}) \over p_{\bf n}({\bf n}) } \, p_{\bf s}({\bf s} | \epsilon).
\label{Lambdaformula1}
\eeq

The formula for $\Lambda({\bf x},\epsilon)$ becomes
more complex when there are unknown signal and/or noise parameters
present.  Suppose the signal distribution depends on some parameters
$\bftheta_s$ in addition to the signal amplitude $\epsilon$, which
themselves are distributed according to a prior 
probability distribution $p_{\theta_s}(\bftheta_s | \epsilon)$.  Then
the signal distribution $p_{\bf s}({\bf s} | \epsilon)$ can be expanded as
\be
p_{\bf s}({\bf s} | \epsilon ) = \int \, d\bftheta_s \, p_{\bf s}({\bf s} |
\epsilon,\bftheta_s) \, p_{\theta_s}(\bftheta_s | \epsilon),
\label{signalPDFexpand}
\eeq
where $p_{\bf s}({\bf s} | \epsilon,\bftheta_s)$ is the distribution
for ${\bf s}$ given both $\epsilon$ and $\bftheta_s$.
Similarly suppose that the noise distribution contains unknown
parameters $\bftheta_n$, whose {\it a priori} distribution 
is $p_{\theta_n}(\bftheta_n)$.  Then the noise distribution can be
expanded as
\be
p_{\bf n}({\bf n}) = \int \, d\bftheta_n \, p_{\bf n}({\bf n} |
\bftheta_n) \, p_{\theta_n}(\bftheta_n).
\label{noisePDFexpand}
\eeq
Inserting the expansions (\ref{signalPDFexpand}) and
(\ref{noisePDFexpand}) into Eq.\ (\ref{Lambdaformula1}) gives the
final expression for the likelihood function:
\begin{eqnarray}
\Lambda({\bf x} , \epsilon)& = &\int d {\bf s} \, \  
{ \int d \bftheta_n p_{\bf n}({\bf x} - {\bf s} | \bftheta_n) \,
p_{\theta_n}(\bftheta_n)  \over 
 \int d \bftheta_n^\prime p_{\bf n}({\bf x}| \bftheta_n^\prime) \,
p_{\theta_n}(\bftheta_n^\prime)} \nonumber \\
\mbox{} &&
\times \int d\bftheta_s \, p_{\bf s}({\bf s} |
\epsilon,\bftheta_s) \, p_{\theta_s}(\bftheta_s | \epsilon). 
\label{Lambdaformula2}
\end{eqnarray}

Equations (\ref{Lambdaformula0}) and (\ref{Lambdaformula2}) are the
foundational equations that we will use throughout this paper to
compute the likelihood ratio $\Lambda({\bf x})$.  A key feature of
this formalism is that the noise parameters $\bftheta_n$ are treated
as unknowns, to be measured from the detector data along with the
gravitational wave signal, rather than being treated as known a
priori.  That feature underlies the elimination of the
auto-correlation terms encountered in paper I in the case of
a stochastic gravitational wave background.

In the next few sections we will revisit several of the signal
detection problems considered in paper I.  In each case, we will show
that $\Lambda({\bf x})$ is, to a good approximation, a monotonic
function of the locally-optimal detection statistic derived in paper
I.  Since a monotonic function of a detection statistic has the same
false alarm versus false dismissal curve as the original statistic,
it follows that in each case the uniquely determined Bayesian statistic
$\Lambda({\bf x})$ is equivalent to the statistic computed in paper
I.  Note, however, that the equivalence only applies at the level of
choosing the detection statistic, and not at the level of specifying
thresholds.  The Bayesian and frequentist
approaches lead to different detection thresholds for a given
specified significance level; see, for example, the discussion in Sec.\
III.C of Ref.\ \cite{powerfilter}.

In deriving the formulae for the likelihood ratio $\Lambda({\bf x})$,
we shall invoke a number of different approximations.  In assessing
the validity of those approximations, we shall be concerned only
with their effect on the false alarm versus false dismissal curve of
the statistic.  In other words, the approximations might be very
inaccurate for computing the value of $\Lambda({\bf x})$, but might
nevertheless be very accurate in the sense that they have only a small
effect on the false alarm versus false dismissal curve.
[We do need to compute accurate numerical values of $\Lambda({\bf x})$,
since we are not concerned here with computing detection thresholds.]
We shall use the notation 
\be
\Lambda_1({\bf x}) \simeq \Lambda_2({\bf x})
\ee
to mean that the false alarm versus false dismissal curves of the
statistics $\Lambda_1({\bf x})$ and $\Lambda_2({\bf x})$ are
approximately the same.

\section{DETERMINISTIC SIGNALS}
\label{s:deterministic}

\subsection{Single detector, white Gaussian noise, known variance}
\label{ss:sdwnkw}

We first treat the simple case where we are looking for a signal
${\bf s}$, in a single detector, whose values in the time domain are
\footnote{The quantity denoted by ${\hat s}_j$ here was
denoted by $s_j$ in paper I.} 
\be
s_j = \epsilon {\hat s}_j.
\ee
We assume that the quantities ${\hat s}_j$ are known and fixed, so
that the only unknown parameter characterizing the signal is its
amplitude $\epsilon$, which can be positive or negative.  Without loss
of generality we can choose the normalization so that
\be
\sum_j |{\hat s}_j|^2 =1.
\label{norm}
\ee
We assume that the detector noise is white and Gaussian with zero mean
and unit variance.  Then, as shown in paper I, the distribution for
the data ${\bf x}$ given $\epsilon$ is
\be
p({\bf x}|\epsilon) = \prod_{j} {1 \over \sqrt{2 \pi}} \exp \left[ - { 1
\over 2} (x_j - \epsilon {\hat s}_j)^2 \right].
\label{pIresult}
\ee
Inserting this formula into Eq.\ (\ref{Lambdaformula}) gives
\be
\Lambda({\bf x},\epsilon) = \exp \left[ \epsilon {\hat \epsilon}({\bf
x}) - \epsilon^2/2 \right],
\label{Lambda0}
\ee
where
\be
{\hat \epsilon}({\bf x}) = \sum_j \, x_j {\hat s}_j
\ee
is the standard matched filtering statistic.  Combining this with Eq.\
(\ref{Lambdaformula0}) gives for the likelihood ratio 
\be
\Lambda({\bf x}) = e^{ {\hat \epsilon}({\bf x})^2/2} \int d\epsilon \,
p^{(0)}(\epsilon) e^{ - [\epsilon - {\hat \epsilon}({\bf x})]^2/2}.
\label{ii}
\ee

Now the quantity $|{\hat \epsilon}({\bf x})|$ is effectively the
signal-to-noise ratio.  Let us assume that we are in the relevant regime where
the signal is detectable with high confidence, so that
\be
\exp \left[ {\hat \epsilon}({\bf x})^2/2 \right] \gg 1.
\label{detectable}
\ee
Let us also assume that the prior distribution $p^{(0)}(\epsilon)$ 
is slowly varying and does not strongly constrain the possible values
of $\epsilon$.  Then, we can approximately evaluate the integral
(\ref{ii}) using the Laplace approximation to obtain
\be
\Lambda({\bf x}) \approx \sqrt{2 \pi} p^{(0)}[ {\hat \epsilon}({\bf
x})] \, \exp \left[ {\hat \epsilon}({\bf x})^2/2 \right].
\label{Lambdaans}
\ee
Finally, we argue that we can neglect the dependence on ${\bf x}$ of the
factor $p^{(0)}[ {\hat \epsilon}({\bf x})]$ in the expression
(\ref{Lambdaans}).  The reason is that the prior distribution
$p^{(0)}(\epsilon)$ is a slowly varying function of $\epsilon$, and so
this factor has a much weaker dependence on ${\bf x}$ than the
exponential factor in the regime (\ref{detectable}).  Therefore,
dropping the factor  
$p^{(0)}[ {\hat \epsilon}({\bf x})]$ will have a
negligible effect on the false alarm versus false dismissal curve of
the statistic.   In this approximation we
see that $\Lambda({\bf x})$ is a monotonic function of the standard
detection statistic $|{\hat \epsilon}({\bf x})|$, 
\be
\Lambda({\bf x}) \simeq \exp \left[ {\hat \epsilon}({\bf x})^2/2
\right],
\label{ans1}
\ee
as claimed
\footnote{In the case where the prior distribution $p^{(0)}(\epsilon)$
restricts 
the sign of $\epsilon$ to be positive, one can similarly show that, in
the relevant regime given by ${\hat \epsilon}({\bf x}) > 0$ together with the
condition (\ref{detectable}), $\Lambda({\bf x})$ is to a good
approximation a monotonic function of the standard detection statistic
$
\Theta[{\hat \epsilon}({\bf x})] \, {\hat \epsilon}({\bf x}),
$
where $\Theta$ is the step function.  Similar comments apply to the
statistics (\ref{ans2}) and (\ref{l4}), .}.

We remark that there is a key technical difference between the
above computation and the corresponding computation in
Sec. II.A. of paper I.  The Bayesian computation presented here
requires expanding the quantity $\ln \Lambda({\bf x},\epsilon)$ to
second order in $\epsilon$ about $\epsilon=0$ [Eq.\ (\ref{Lambda0})
above], whereas in 
paper I it sufficed to compute $\ln \Lambda({\bf x},\epsilon)$ to
linear order in $\epsilon$ [Eqs. (2.3) and (2.5) of paper I].  This
difference is a common feature of all of our subsequent computations.

\subsection{Single detector, white Gaussian noise, unknown variance}
\label{ss:sdwnkwu}

We now add one additional complication to the analysis, by taking the
noise variance to be an unknown constant $\sigma$.  We define an inner
product $\left< \, , \, \right>$ on the space of 
signals by
\be
\left< {\bf x} , {\bf y} \right> \equiv \sum_j x_j y_j.
\ee
Note that this is {\it not} the standard inner product used in
discussions of matched filtering, which incorporates a weighting 
factor of $\sigma^{-2}$.  We define the statistic 
${\hat \sigma}({\bf x})$ by
\be
{\hat \sigma}({\bf x})^2 \equiv {1 \over N} \left< {\bf x} , {\bf x}
\right>,
\label{hatsigmadef}
\ee
where $N$ is the number of data points; this is the conventional
estimator of $\sigma$.  We also define the quantity $\rho$ by
\be
\rho = { \epsilon \over \sigma},
\ee
which from the normalization condition (\ref{norm}) is the
conventional signal to noise ratio.  The corresponding  
estimator is
\be
{\hat \rho}({\bf x}) = {\epsilon \over {\hat \sigma}({\bf x})}.
\ee
The conventional matched filtering statistic is
\be
{1 \over \sigma} \left< {\bf x} , {\hat s} \right>,
\ee
and if we replace the noise variance by its estimator ${\hat
\sigma}({\bf x})$ we obtain the statistic
\be
{\hat \rho}_1({\bf x}) \equiv {1 \over {\hat \sigma}({\bf x})} \left<
{\bf x} , {\hat s} \right>.
\label{hatrho1def}
\ee
We shall show below that the likelihood ratio $\Lambda({\bf x})$ is to
a good approximation equivalent to the conventional statistic
(\ref{hatrho1def}).

The noise
distribution given $\sigma$ is taken to be
\be
p_{\bf n}({\bf n}|\sigma) = \prod_j \, {1 \over \sqrt{2 \pi} \sigma}
\exp \left[ - {n_j^2 \over 2 \sigma^2} \right].
\label{n1}
\ee
The full noise distribution is [cf.\ Eq.\ (\ref{noisePDFexpand}) above]
\be
p_{\bf n}({\bf n}) = \int_0^\infty d\sigma \, p_\sigma(\sigma) \,
p_{\bf n}({\bf n}|\sigma),
\label{n2}
\ee
where $p_\sigma(\sigma)$ is the prior probability distribution for $\sigma$.
Inserting Eq.\ (\ref{n2}) into Eq.\ (\ref{n1}) and using the definition
(\ref{hatsigmadef}) we obtain
\be
p_{\bf n}({\bf n}) = \int_0^\infty d\sigma \, p_\sigma(\sigma) \,
\exp \left[ - N \Xi(\sigma) /2 \right],
\label{n3}
\ee
where
\be
\Xi(\sigma) = \ln ( 2 \pi \sigma^2) + { {\hat \sigma}({\bf n})^2 \over
\sigma^2}.
\ee
We can approximately evaluate the integral (\ref{n3}) in the limit
where $N$ is large.  The function $\Xi(\sigma)$ can be expanded about
its local minimum at $\sigma = {\hat \sigma}$ as
\be
\Xi(\sigma) = 1 + \ln( 2 \pi {\hat \sigma}^2) + {2 \over {\hat
\sigma}^2} (\sigma - {\hat \sigma})^2 + O[ (\sigma-{\hat \sigma})^3].
\ee
Using this expansion we obtain
\begin{eqnarray}
p_{\bf n}({\bf n}) &=& \sqrt{\pi \over N} (2 \pi e)^{-{N \over 2}} 
p_\sigma[ {\hat \sigma}({\bf n}) ] \, \, {\hat \sigma}({\bf
n})^{-(N-1)} \nonumber \\
\mbox{} && \times \left[ 1 + O\left({ 1 \over \sqrt{N}}\right) \right].
\label{pnfinal}
\end{eqnarray}
We assume that $p_\sigma(\sigma)$ is slowly varying, and so have
neglected in Eq. (\ref{pnfinal}) a fractional error of order the
fractional change in $p_{\bf n}$ over a interval of width ${\hat
\sigma}/\sqrt{N}$.

We next insert the formula (\ref{pnfinal}) for the noise distribution
into the expression (\ref{Lambdaformula1}) for the likelihood ratio,
using 
\be
p_{\bf s}({\bf s}|\epsilon) = \delta^N( {\bf s} - \epsilon {\hat {\bf s}}).
\label{signalknown}
\ee
The result is 
\begin{eqnarray}
&& \Lambda({\bf x}) = \int_0^\infty d\epsilon p^{(0)}(\epsilon) \, {
p_\sigma[ {\hat \sigma}({\bf x} - \epsilon {\hat {\bf s}})] \over
p_\sigma[ {\hat \sigma}({\bf x})] }
\left[{  {\hat \sigma}({\bf x} - \epsilon {\hat {\bf s}}) \over 
{\hat \sigma}({\bf x}) } \right]^{-(N-1)} \\
&&= \int_0^\infty d\epsilon p^{(0)}(\epsilon) \, {
p_\sigma[ {\hat \sigma}({\bf x} - \epsilon {\hat {\bf s}})] \over
p_\sigma[ {\hat \sigma}({\bf x})] } \nonumber \\
\mbox{} && \times \exp \left[ -{N-1 \over 2} \ln \left\{ 1 - 2 \epsilon 
{ \left< {\bf x} , {\hat {\bf s}} \right> \over \left< {\bf x}, {\bf x}
\right> } + 
{ \epsilon^2 \over \left< {\bf
x}, {\bf x} \right>^2 } \right\} \right].
\label{Lambdaformula3}
\end{eqnarray}
To obtain the second line we used Eqs.\ (\ref{norm}) and (\ref{hatsigmadef}).
Expanding the logarithm to second order in
$\epsilon$, we can re-express this as
\begin{eqnarray}
\Lambda({\bf x}) &=& \int_0^\infty d\epsilon p^{(0)}(\epsilon) \, {
p_\sigma[ {\hat \sigma}({\bf x} - \epsilon {\hat {\bf s}})] \over
p_\sigma[ {\hat \sigma}({\bf x})] } \nonumber \\
\mbox{} &&
\times \exp \left[ {\hat a} \epsilon - {\hat b} \epsilon^2 + O(\epsilon^3)
\right],
\label{Lambdaformula3a}
\end{eqnarray}
where
\be
{\hat a} = (N-1) { \left< {\bf x} , {\hat {\bf s}} \right> \over 
\left< {\bf x} , {\hat {\bf x}} \right> }
\label{ahatdef}
\ee
and
\be
{\hat b} = {1 \over 2} (N-1) \left[ 
{ 1 \over 
\left< {\bf x} , {\hat {\bf x}} \right> } - 2
{ \left< {\bf x} , {\hat {\bf s}} \right>^2 \over 
\left< {\bf x} , {\hat {\bf x}} \right>^2 } \right].
\label{bhatdef}
\ee

Before proceeding further with the computation of the likelihood ratio,
we clarify the domain of validity of the 
weak signal expansion (expansion in powers of $\epsilon$) used in
going from Eq.\
(\ref{Lambdaformula3}) to Eq.\ (\ref{Lambdaformula3a}).   We will
estimate the 
expected sizes and the scale of statistical fluctuations in the two
terms appearing in the argument of the logarithm in Eq.\
(\ref{Lambdaformula3}); the expansion will be good when both of these 
terms and their fluctuations are small compared to unity.  
For the purpose of making these estimates we can identify $\sigma$ and
${\hat \sigma}$, and $\rho$, ${\hat \rho}$, and ${\hat \rho}_1$.

We can compute the expected value and variance of the statistic
$\left< {\bf x} , {\bf x} \right>$ using Eqs.\ (\ref{additivenoise}),
(\ref{epsilondef}) and (\ref{n1}), which gives
\be
\left< {\bf x} , {\bf x} \right> \sim (N \sigma^2 + \epsilon^2) \pm
\sqrt{2 N \sigma^4 + 4 \epsilon^2 \sigma^2}.
\ee
The notation here is that the first term gives the expected value, and
the second term gives an estimate of the statistical fluctuations.
We can rewrite this formula in terms of the signal-to-noise ratio
$\rho = \epsilon/\sigma$ as
\be
\left< {\bf x} , {\bf x} \right> \sim N \sigma^2 \left[ (1 + {\rho^2
\over N}) \pm
\sqrt{{2 \over N} + 4 {\rho^2 \over N^2}} \right].
\label{ps}
\ee
Similarly we have
\be
\left< {\bf x} , {\hat {\bf s}} \right> \sim \epsilon \pm \sigma.
\label{mfs}
\ee
We assume that $N \gg 1$ and that $\rho \agt 1$.  We now consider two
different cases:  

\begin{itemize}

\item When $\rho^2 / N \ll 1$, the
fluctuations in $\left< {\bf x} , {\bf x} \right>$ are small compared
to the expected value, and we have from Eq.\ (\ref{ps}) that 
$\left< {\bf x} , {\bf x} \right> \sim N \sigma^2 \pm \sqrt{N}
\sigma^2$.  Using this 
together with Eq.\ (\ref{mfs}) shows that the first term in the
argument of the logarithm in Eq.\ (\ref{Lambdaformula3}) is
\be
\epsilon { \left< {\bf x} , {\hat {\bf s}} \right> \over \left< {\bf
x}, {\bf x} \right> } \sim {\rho^2 \over N} \pm {\rho \over N} \ll 1,
\ee
and similarly the second term is 
\be
 { \epsilon^2 \over
\left< {\bf x}, {\bf x} \right> } \sim {\rho^2 \over N} \pm {\rho^2
\over N^{3/2}} \ll 1.
\ee
Thus the approximation is good in this regime.

\item
When $\rho^2 /N \gg 1$, a similar computation gives that $\langle {\bf
x} , {\bf x} \rangle \sim \rho^2 \sigma^2 \pm \rho \sigma^2$.  The 
the first term in the argument of the logarithm in Eq.\
(\ref{Lambdaformula3}) now scales as
\be
\epsilon { \left< {\bf x} , {\hat {\bf s}} \right> \over \left< {\bf
x}, {\bf x} \right> } \sim 1 \pm {1 \over \rho}
\ee
and similarly the second term scales as 
\be
{\epsilon^2  \over
\left< {\bf x}, {\bf x} \right> } \sim 1 \pm {1 \over \rho}.
\ee
Thus, the approximation breaks down in this regime.

\end{itemize}

We now return to computing the likelihood ratio $\Lambda({\bf x})$.
We can approximately evaluate the integral (\ref{Lambdaformula3a})
using the Laplace approximation to
obtain  
\be
\Lambda({\bf x}) \approx
p^{(0)}({\hat \epsilon}) \, {
p_\sigma[ {\hat \sigma}({\bf x} - {\hat \epsilon} {\hat {\bf s}})] \over
p_\sigma[ {\hat \sigma}({\bf x})] } \sqrt{ {\hat a} \pi \over {\hat b}}
\exp \left[ {{\hat a}^2 \over 4 {\hat b}}\right],
\label{Lambdaformula3b}
\ee
where 
\be
{\hat \epsilon}({\bf x}) = { {\hat a}({\bf x}) \over 2 {\hat b}({\bf
x})}.
\label{ans2a}
\ee
This approximation will be good whenever the 
exponential factor in Eq.\ (\ref{Lambdaformula3b}) is large, which it
will be in the regime where the signal is detectable (see below), and
when the prior probability distribution $p^{(0)}(\epsilon)$ is slowly
varying.  Using Eqs.\ (\ref{hatsigmadef}), (\ref{hatrho1def}),
(\ref{ahatdef}) and (\ref{bhatdef}), we can 
write the exponential factor as
\be
\exp \left[{N-1 \over 2} g({\hat \rho}_1^2/N) \right],
\ee
where the function $g$ is given by $g(x) = x / (1 - 2 x)$.
Since we are in the regime $\rho^2/N \ll 1$, the argument of the
function $g$ is small, and we can replace $g({\hat \rho}_1^2/N)$ by
$\rho_1^2/N$.  This gives, using $N \gg 1$,
\be
\Lambda({\bf x}) \approx
p^{(0)}({\hat \epsilon}) \, {
p_\sigma[ {\hat \sigma}({\bf x} - {\hat \epsilon} {\hat {\bf s}})] \over
p_\sigma[ {\hat \sigma}({\bf x})] } \sqrt{ {\hat a} \pi \over {\hat b}}
\exp \left[ {1 \over 2} {\hat \rho}_1({\bf x})^2 \right].
\label{Lambdaformula3c}
\ee
Finally, we argue as before that in the regime $\exp[ \rho^2/2] \gg 1$
where the signal is detectable, the dependence on ${\bf x}$ of all the
other factors in Eq. (\ref{Lambdaformula3c}) can be neglected in
comparison to the exponential factor, assuming that the prior
distributions are slowly varying.  This gives
\be
\Lambda({\bf x}) \simeq \exp[ {\hat \rho}_1({\bf x})^2/2 ],
\label{ans2}
\ee
as claimed.

Our final answer (\ref{ans2}) is essentially the same as the answer
(\ref{ans1}) obtained when 
the noise variance $\sigma$ is assumed to be known.  Therefore,
treating $\sigma$ as an unknown parameter rather than as a fixed,
known parameter does not make much difference in this case.  However,
we will see below for the case of stochastic signals that treating the
properties of the noise distribution as unknowns does have a
significant effect on the analysis, and that the correct answer is
obtained only when the those properties are treated as unknowns.

We end this subsection by recapitulating the various approximations
and assumptions we have invoked:

\begin{itemize}

\item The large $N$ approximation $N \gg 1$.

\item The assumption that we are in the regime where the signal is
detectable, $\exp( \rho^2/2) \gg 1$.  
This is necessary for evaluating the integral over $\epsilon$ to
obtain Eq.\ (\ref{Lambdaformula3b}), and also for the validity in
neglecting the prefactors in deriving Eq.\ (\ref{ans2}).
From a practical point of view the assumption $\exp( \rho^2/2) \gg 1$
is not a serious restriction, as it does not matter how our
statistics perform in the regime $\exp( \rho^2/2 ) \sim 1$ where
signals are not detectable.

\item The assumption that the prior probability distributions
$p^{(0)}(\epsilon)$ and $p_\sigma(\sigma)$ are slowly varying.

\item We have clarified the ``weak signal'' assumption of paper I; it
is the assumption is that the signal-to-noise ratio 
squared per data point is small, $\rho^2/N \ll 1$.  This 
requirement ensures that the presence of the signal does not significantly
bias the estimate (\ref{hatsigmadef}) of the noise variance.
In practice we can always choose segments of data large enough to
satisfy this assumption.  
\end{itemize}

\subsection{Single detector, white non-Gaussian noise}
\label{ss:color}

We now turn to the case where the noise has a known, non-Gaussian
distribution.  In this subsection we  
follow Sec. II.A. of paper I, and assume 
that the noise samples in the time
domain are statistically independent but identically distributed, with
a known distribution.  We can write the noise probability distribution
as 
\be
p_{\bf n}({\bf n}) = \prod_j e^{- f(n_j)}.
\ee
We assume that the probability distribution $e^{-f(x)}$ has a central
Gaussian region $|x| < x_b$ in which
\be
f(x) = {x^2 \over 2 \sigma^2},
\ee
which contains most of the probability, and
a tail region $|x| \ge x_b$ containing a total probability $p_{\rm tail}$
with $p_{\rm tail} \ll 1$.

As before, the signal is assumed to be known up to an overall
amplitude parameter.  From Eqs.\ (\ref{finn}) and (\ref{signalknown})
we obtain the following 
modified version of Eq.\ (\ref{pIresult}):
\be
p({\bf x}|\epsilon) = \prod_{j} \exp \left[ - f (x_j - \epsilon {\hat
s}_j) \right],
\ee
and inserting this into Eqs.\ (\ref{Lambdaformula0}) and
(\ref{Lambdaformula}) gives 
\be 
\Lambda({\bf x}) = \int d\epsilon p^{(0)}(\epsilon) \, \prod_j \exp
\left[ -f(x_j - \epsilon {\hat s}_j) + f(x_j) \right].
\ee
Expanding to second order in $\epsilon$ gives
\be
\Lambda({\bf x}) = \int d\epsilon \, p^{(0)}(\epsilon) \exp \left[
{\hat a}({\bf x}) \epsilon - {\hat b}({\bf x}) \epsilon^2 +
O(\epsilon^3) \right],
\label{ans3}
\ee
where 
\be
{\hat a}({\bf x}) = \sum_j f'(x_j) {\hat s}_j
\label{ahatdef1}
\ee
and
\be
{\hat b}({\bf x}) = {1 \over 2} \sum_j f^{\prime\prime}(x_j) {\hat
s}_j^2.
\label{bhatdef1}
\ee
Evaluating the integral over $\epsilon$ using the same types of
arguments as in Sec.\ \ref{ss:sdwnkwu} gives
\be
\Lambda({\bf x}) \simeq \exp \left[ {{\hat a}({\bf x})^2 \over 2 {\hat
b}({\bf x})} \right].
\label{l4}
\ee

Now the statistic ${\hat a}({\bf x})$ is the locally optimal statistic
computed in paper I [Eq.\ (2.9) of paper I].  Therefore it remains to
show that we can neglect the ${\bf x}$ dependence of the factor ${\hat
b}({\bf x})$ in the argument of the exponential in Eq.\ (\ref{l4}).

We can split the sum (\ref{bhatdef1}) into
contributions from the Gaussian region and from the tail.  Using the
fact that $f^{\prime\prime}(x) = 1/\sigma^2$ in the Gaussian region,
we obtain
\be
{\hat b}({\bf x}) =
{1 \over 2} \sum_{x_j < x_b} {{\hat s}_j^2 \over \sigma^2}
+ {1 \over 2} \sum_{x_j \ge x_b}  f^{\prime\prime}(x_j) {\hat s}_j^2.
\ee
Most of the values of $x_j$ will fall in the central Gaussian region,
since $x_j = n_j + \epsilon {\hat s}_j$, and $\epsilon {\hat s}_j \ll
\sigma$ for each individual $j$ \footnote{This follows from the
relation $\rho^2/N = {\hat s}_{\rm rms}^2 \epsilon^2 / \sigma^2$, where
${\hat s}_{\rm rms}$ is the rms average of the quantities ${\hat
s}_j$, and from the assumption $\rho^2/N \ll 1$.}.  Since $\sum_j {\hat
s}_j^2=1$, the 
first term gives $1/(2 \sigma^2)$, up to fractional corrections of
order $p_{\rm tail}$.  Similarly the second term will be bounded above
by $\sim p_{\rm tail} / (2 \sigma^2)$, since $f^{\prime\prime}(x)$ will
be smaller in the tails than in the central Gaussian region.  We
conclude that
\be
{\hat b}({\bf x}) = {1 \over 2 \sigma^2} \left[ 1 + O(p_{\rm tail})
\right].
\ee
It follows in particular that the ${\bf x}$-dependent fluctuations in
${\hat b}({\bf x})$ are smaller than its expected value by a factor of
$p_{\rm tail} \ll 1$, and therefore we can neglect the ${\bf x}$
dependence of ${\hat b}({\bf x})$ in Eq.\ (\ref{l4}), as required.

\medskip
\subsection{Single detector, colored non-Gaussian noise}
\label{ss:sdcnkw}

We next consider the model of colored, non-Gaussian noise of Sec.\
II. B. of paper I, where each frequency bin is assumed to be
statistically independent.  This is given by
\be
p_{\bf n}({\bf n}) = \prod_{k=1}^{[(N-1)/2]} {2 \over \pi P_k}  
  \exp \left[ -2 g_k \left({ |\tilde n_k|^2 \over P_k}\right) \right],
\ee
where the volume element is understood to be
\be
\prod_{k=1}^{[(N-1)/2]}   d( {\rm Re} \, {\tilde n}_k) \, d( {\rm Im}
\, {\tilde n}_k).
\ee
Here 
\be
{\tilde n}_k = {1 \over \sqrt{N}} \sum_{j} e^{2 \pi i j k/N} \, n_j
\ee
are the components of the discrete Fourier transform of the time
domain samples $n_j$.  The quantities $P_k$ describe the noise
spectrum.  For each frequency bin $k$, the function $g_k(x)$ is 
arbitrary except for the normalization conditions 
\be
\int_0^\infty dx e^{-g_k(x)} = \int_0^\infty dx \, x e^{-g_k(x)} = 1,
\ee
and the requirement that $g_k(x) = x$ in a central Gaussian region
containing most of the probability.

By paralleling the analysis of Sec.\ \ref{ss:color}, we again arrive at the
formulae (\ref{ans3}) and (\ref{l4}), where now the statistics ${\hat
a}({\bf x})$ and ${\hat b}({\bf x})$ are given by
\be
{\hat a}({\bf x}) = 4 \sum_k g_k^\prime \left( { |{\tilde x}_k|^2
\over P_k } \right) \, { {\rm Re} ( {\tilde x}_k^* {\tilde s}_k )
\over P_k}
\label{ahatdef2}
\ee
and
\begin{eqnarray}
{\hat b}({\bf x}) &=& 
2 \sum_k g_k^\prime \left( { |{\tilde x}_k|^2
\over P_k } \right) \, {  |{\tilde s}_k |^2
\over P_k} \nonumber \\
\mbox{} && + \, 4 \sum_k g_k^{\prime\prime} \left( { |{\tilde x}_k|^2
\over P_k } \right) \, { \left[{\rm Re} ( {\tilde x}_k^* {\tilde s}_k
) \right]^2 \over P_k^2}.
\label{bhatdef2}
\end{eqnarray}
As before, the statistic ${\hat a}({\bf x})$ coincides with the locally
optimal statistic derived in paper I [Eq.\ (2.21) of paper I], and it
suffices to show that the ${\bf x}$ dependence of the factor ${\hat
b}({\bf x})$ can be neglected in Eq.\ (\ref{l4}).
We evaluate the sums in Eq.\ (\ref{bhatdef2}) by splitting them into
Gaussian and tail contributions as before.  Since $g_k^\prime(x) =1$
in the Gaussian region, the first term in Eq.\ (\ref{bhatdef2}) yields
$2 \sum_k | {\tilde s}_k|^2 / P_k [ 1 + O(p_{\rm tail})]$.  Also the
second term is proportional to $p_{\rm tail}$ since
$g_k^{\prime\prime}(x)$ vanishes in the Gaussian region.
Thus we obtain
\be
{\hat b}({\bf x}) = \left[2 \sum_k { | {\tilde s}_k |^2 \over P_k}
\right]\, \bigg[ 1 + O(p_{\rm tail}) \bigg],
\ee
and the rest of the argument follows as before.

\subsection{Signals with unknown parameters}

We now generalize the analysis of the preceding subsections by
allowing the signals to depend on additional parameters other than the
overall amplitude parameter $\epsilon$.  We write
\be
{\bf s} = \epsilon {\hat {\bf s}}(\bftheta_s),
\ee
where the signal parameters $\bftheta_s$ are distributed according to
the distribution $p_{\theta_s}(\bftheta_s | \epsilon)$.  Then from
Eq.\ (\ref{Lambdaformula2}) we can write
\be
\Lambda({\bf x}) = \int d\epsilon \int d\bftheta_s p^{(0)}(\epsilon)
p_{\theta_s}(\bftheta_s|\epsilon) \Lambda({\bf x},\epsilon,\bftheta_s),
\ee
where $\Lambda({\bf x},\epsilon,\bftheta_s)$ is given by Eq.\
(\ref{finn}) with $p_{\bf s}({\bf s}|\epsilon)$ replaced by 
$p_{\bf s}({\bf s}|\epsilon,\bftheta_s)$.
In the regime where the signal is detectable, we can 
repeat the arguments of the preceding subsections to approximately
evaluate the integrals as
\be
\Lambda({\bf x}) \simeq \max_{\bftheta_s} \ \max_\epsilon \Lambda({\bf
x},\epsilon,\bftheta_s).
\ee
Thus the result is to take the statistics previously derived
and to maximize over the signal parameters.  
Such a maximization is the standard thing to do for linear matched
filtering; the above argument indicates that it is also the
appropriate procedure for the more general class of locally optimal
statistics.

\section{STOCHASTIC SIGNALS}
\label{s:stochastic}

The standard method of detecting a stochastic background is to compute
a cross-correlation between two different instruments
\cite{crosscorr}; see Ref.\ \cite{allen96} for a detailed description.
In Sec.\ \ref{ss:stochas1} below we compute the   
likelihood ratio  
$\Lambda({\bf x})$ for the simplest case of a white stochastic signal,
and of two co-aligned detectors
with white Gaussian noise, where the noise variance is assumed to be
known.  For this case we do not recover the standard cross-correlation
statistic, but instead we obtain a statistic with extra auto-correlation
terms.  That statistic was first derived and has been investigated in
detail in a more general context by Finn and Romano \cite{FinnRomano}.
We then argue that it is unrealistic to take the noise variances to be
known parameters.  In Sec.\ \ref{ss:stochas2} we show that, when the noise
variances are taken to be unknowns to be determined from the data, then
the likelihood ratio $\Lambda({\bf x})$ is to a good approximation
equivalent to the standard cross-correlation statistic.  This
computation is again in the simple context of coincident aligned detectors with
white noise.  The computation of Sec.\ \ref{ss:stochas2} is a
simplified version of the computation in Appendix A of Ref.\
\cite{flanagan93}.   

We then turn to non-Gaussian noise models.  In paper I, we derived a
generalized cross-correlation (GCC) statistic appropriate for
non-Gaussian noise, which is a modification of the 
standard cross-correlation statistic.  In Sec.\ \ref{ss:stochas3}
below we re-derive that statistic using the Bayesian approach.

\subsection{Two coincident co-aligned detectors, white Gaussian noise,
known variances}
\label{ss:stochas1}

The output of the pair of detectors is ${\bf x} = ({\bf x}_1 , {\bf
x}_2)$,
where
\be
{\bf x}_1 = {\bf n}_1 + {\bf s}
\ee
is the output of the first detector, and
\be
{\bf x}_2 = {\bf n}_2 + {\bf s}
\ee
is the output of the second.  We assume, for simplicity, that the noise in each
detector is white and Gaussian with unit variance:
\be
p_{{\bf n}_1}({\bf n}_1) = \prod_j {1 \over \sqrt{2 \pi} }
\exp \left[ - {n_{1j}^2 \over 2} \right],
\label{stochasnoise}
\ee
with a similar equation for the second detector.  We assume that the
stochastic background signal is also white and Gaussian with variance
$\epsilon$: 
\be
p_{\bf s}({\bf s}|\epsilon) = \prod_j {1 \over \sqrt{2 \pi \epsilon}}
\exp \left[ - {s_{j}^2 \over 2 \epsilon} \right],
\label{stochassignal}
\ee
where $\epsilon \ge 0$.  As in earlier sections, $\epsilon$
parameterizes the signal strength, and we will be using a weak signal
expansion of expanding in powers of $\epsilon$ about $\epsilon=0$.

By inserting the distributions (\ref{stochasnoise}) and
(\ref{stochassignal}) into the formulae (\ref{finn}) and
(\ref{noisePDFexpand}) we obtain  
\be
p({\bf x}|\epsilon) = \exp \left[ -{N \over 2} \Xi(\epsilon, {\bf x})
\right],
\label{nsd}
\ee
where
\begin{eqnarray}
\Xi(\epsilon,{\bf x}) &=& 2 \ln(2 \pi) + \ln \left[ 1
 + 2 \epsilon \right] \nonumber \\
\mbox{} && + { (1 + \epsilon) {\hat \sigma}_2^2 + (1
+ \epsilon) {\hat \sigma}_1^2 - 2 \epsilon {\hat \epsilon} \over
1 + 2 \epsilon},  
\label{Xidef}
\end{eqnarray}
and where
\be
{\hat \sigma}_1^2 = {1 \over N} \sum_j x_{1j}^2,
\label{hatsigma1def}
\ee
\be
{\hat \sigma}_2^2 = {1 \over N} \sum_j x_{2j}^2,
\label{hatsigma2def}
\ee
and
\be
{\hat \epsilon} = {1 \over N} \sum_j x_{1j} x_{2j}.
\label{hatepsilondef}
\ee
The statistic ${\hat \epsilon}$ is the standard cross-correlation
statistic. 

Before proceeding further we discuss the validity of the weak signal
approximation in the context of a stochastic background signal.
Suppose that a stochastic background is present and just
barely detectable by cross-correlating between the two detectors.  Then we
have ${\hat \epsilon} \sim 1/\sqrt{N}$.  In the context of ground
based detectors like LIGO, when this analysis is generalized to
colored noise, $N$ is replaced by the product $T \Delta f$, where $T$
is the observation time and $\Delta f$ is the effective bandwidth in
the usual formula for signal-to-noise ratio [Eq.\ (1.2) of Ref.\
\cite{flanagan93}].  Using the estimates $T 
\sim 1/3 $ year and 
$\Delta f \sim 50 \, {\rm Hz}$ we find
\beq
{\hat \epsilon} \sim {1 \over \sqrt{T \Delta f}} \sim 10^{-4}.
\eeq
Therefore in all our analyses it will be sufficient to work to first
order in ${\hat \varepsilon}$.  The approximation would only break down
if ${\hat \epsilon}\sim 1$, that is, if the stochastic background could be seen
in a single detector, which is thought to be very unlikely.

We define the statistics ${\hat d}_1$ and ${\hat d}_2$ by
\be
{\hat d}_1({\bf x}) = {\hat \sigma}_1({\bf x})^2 - 1 - {\hat
\epsilon}({\bf x})
\label{d1def}
\ee
and
\be
{\hat d}_2({\bf x}) = {\hat \sigma}_2({\bf x})^2 - 1 - {\hat
\epsilon}({\bf x}).
\label{d2def}
\ee
Now $\left< {\hat \sigma}_1^2 \right> = 1 + \epsilon$, so the quantities
${\hat d}_1$ and ${\hat d}_2$ will be small;
\be
|{\hat d}_{1,2}({\bf x})| \alt O({1 /\sqrt{N}}) + O({\hat \epsilon}).
\ee

We now insert Eqs.\ (\ref{d1def}) and (\ref{d2def}) into the formula
(\ref{Xidef}) for the function $\Xi$.  We expand to second order in
$\epsilon$, ${\hat \epsilon}$, ${\hat d}_1$, and ${\hat d}_2$,
treating these quantities as formally all of the same order.
We then insert the result into Eqs.\ (\ref{Lambdaformula0}),
(\ref{Lambdaformula}), and (\ref{nsd}), which yields
\be
\Lambda({\bf x}) = \int d\epsilon p^{(0)}(\epsilon) \exp \left[ -N
(\epsilon - {\hat \epsilon}_b)^2 + N {\hat \epsilon}_b^2 \right],
\ee
where 
\begin{eqnarray}
{\hat \epsilon}_b({\bf x}) &=& {\hat \epsilon}({\bf x}) + {1 \over 4}
\left[ {\hat d}_1({\bf x}) + {\hat d}_2({\bf x}) \right] \nonumber \\
\mbox{} &=& {1 \over 2} {\hat \epsilon}({\bf x}) + {1 \over 4} \left[
{\hat \sigma}_1({\bf x})^2 -1 \right]
\nonumber \\ \mbox{} && 
+ {1 \over 4} \left[
{\hat \sigma}_2({\bf x})^2 -1 \right].
\label{hatepsilonbdef}
\end{eqnarray}
Evaluating the integral over $\epsilon$ using the same types of
arguments as in previous sections gives, using $p^{(0)}(\epsilon)=0$
for $\epsilon < 0$, 
\be
\Lambda({\bf x}) \approx \Theta({\hat \epsilon}_b) \sqrt{\pi \over N}
p^{(0)}({\hat \epsilon}_b) \exp[ N {\hat \epsilon}_b^2].
\ee
Here $\Theta$ is the step function.  If follows by the same type of
arguments as before that $\Lambda({\bf x}) \simeq \Theta({\hat
\epsilon}_b) \exp[ N {\hat \epsilon}_b^2]$.

Our final result identifies the statistic ${\hat \epsilon}_b$ as the
optimal detection statistic; see Ref.\ \cite{FinnRomano} for a more
general version of this statistic.  From Eq.\ (\ref{hatepsilonbdef})
this statistic is not the standard cross-correlation statistic ${\hat
\epsilon}$, but instead contains the auto-correlation terms ${\hat
\sigma}_1^2-1$ and ${\hat \sigma}_2^2-1$.  The interpretation of these terms is that it
is possible, under the assumptions of this subsection, to measure the
stochastic background signal with just one detector.  If the
detector's noise variance is known, then one can just measure the 
variance of the detector's output and subtract the known noise
variance to reveal the stochastic background contribution.  

Of course, in reality, the noise in detectors is not known {\it
priori}, and is measured from the data.  In particular, there is no
way that the detectors noise can be known beforehand to a fractional
accuracy of $10^{-4}$.  Therefore we have to generalize the preceding
analysis by allowing the noise variances to be unknown parameters.

\subsection{Two coincident co-aligned detectors, white Gaussian noise,
unknown variances}
\label{ss:stochas2}

We assume that the noise in each
detector is white and Gaussian with variances $\sigma_1$ and
$\sigma_2$.  We replace Eq.\ (\ref{stochasnoise}) with
\be
p_{{\bf n}_1}({\bf n}_1|\sigma_1) = \prod_j {1 \over \sqrt{2 \pi} \sigma_1}
\exp \left[ - {n_{1j}^2 \over 2 \sigma_1^2} \right],
\label{stochasnoise1}
\ee
with a similar equation for the second detector.  The prior
distribution for the parameters $\sigma_1$, $\sigma_2$ will be written
as $p_\sigma(\sigma_1,\sigma_2)$.  
By inserting the distributions (\ref{stochasnoise1}) and
(\ref{stochassignal}) into the formulae (\ref{finn}) and
(\ref{noisePDFexpand}) we obtain 
\be
p({\bf x}|\epsilon) = \int_0^\infty d\sigma_1 \int_0^\infty d\sigma_2
p_\sigma(\sigma_1,\sigma_2) \exp \left[ -{N \over 2} \Xi(\epsilon,
\sigma_1,\sigma_2) \right],
\label{pp}
\ee
where the function $\Xi$ is now given by
\begin{eqnarray}
\Xi(\epsilon,\sigma_1,\sigma_2) &&= 2 \ln(2 \pi) + \ln \left[ \sigma_1^2
\sigma_2^2 + \epsilon (\sigma_1^2 + \sigma_2^2) \right] \nonumber \\
\mbox{} && + { (\sigma_1^2 + \epsilon) {\hat \sigma}_2^2 + (\sigma_2^2
+ \epsilon) {\hat \sigma}_1^2 - 2 \epsilon {\hat \epsilon} \over
\sigma_1^2 \sigma_2^2 + \epsilon (\sigma_1^2 + \sigma_2^2) }.  
\label{Xidefnew}
\end{eqnarray}

To evaluate the integral over $\sigma_1$ and $\sigma_2$ in Eq.\
(\ref{pp}), we make a change of variables to variables $f_1$, $f_2$
defined by 
\be
\sigma_1^2 + \epsilon = f_1^2 {\hat \sigma}_1^2
\label{f1def}
\ee
and
\be
\sigma_2^2 + \epsilon = f_2^2 {\hat \sigma}_2^2
\ee
We also define the rescaled variables
\be
\alpha = { \epsilon \over {\hat \sigma}_1 {\hat \sigma}_2 }, \ \ \ \  \
{\hat \alpha} = { {\hat \epsilon} \over {\hat \sigma}_1 {\hat
\sigma}_2 }.
\label{alphadef1}
\ee
We expand $\Xi$ to second order around its local minimum at $f_1 = f_2
=1$, $\alpha = {\hat \alpha}$:
\begin{eqnarray}
\Xi &=& 2 \ln(2 \pi {\hat \sigma}_1 {\hat \sigma}_2) + 2 + \ln (1 -
{\hat \alpha}^2 ) \nonumber \\
\mbox{} && + {1 + {\hat \alpha}^2 \over (1 - {\hat \alpha}^2)^2}
\Delta \alpha^2 -  { 4 {\hat \alpha} \over (1 - {\hat \alpha}^2)^2}
\Delta \alpha
(\Delta f_1 + \Delta f_2) \nonumber \\
\mbox{} && + {2 \over (1 - {\hat \alpha}^2)^2} ( \Delta f_1^2 + \Delta
f_2^2 + 2
{\hat \alpha}^2 \Delta f_1^2 \Delta f_2^2).
\end{eqnarray}
Here $\Delta f_1 = f_1-1$, $\Delta f_2 = f_2-1$, and $\Delta \alpha =
\alpha - {\hat \alpha}$.  At fixed $\alpha$, $\Xi$ is minimized at
\be
f_1 = f_2 = 1 + { {\hat \alpha} \over (1 + {\hat \alpha}^2)}
(\alpha - {\hat \alpha}).
\label{peak}
\ee
We now perform the Gaussian integral
over $\sigma_1$, $\sigma_2$ or $f_1$, $f_2$, which gives
\begin{eqnarray}
p({\bf x}|\epsilon) &=& {p_\sigma[{\hat \sigma}_{1b}(\epsilon),{\hat
\sigma}_{2b}(\epsilon)] \over (2 \pi {\hat \sigma}_1 {\hat
\sigma}_2)^{N-1} }{ (1 - {\hat \alpha}^2)^{3/2} \over 4 (1 +
{\hat \alpha}^2)^{1/2}} {\cal J}(\epsilon) \nonumber \\
\mbox{} && \times \exp \left[ - {N \over 2 (1 + {\hat \alpha}^2)}
(\alpha - {\hat \alpha})^2 \right].
\label{ans4}
\end{eqnarray}
Here ${\hat \sigma}_{1b}(\epsilon)$ is the value of $\sigma_1$ at the
peak (\ref{peak}) of the integrand, given from Eqs.\ (\ref{f1def}) and
(\ref{peak}) by
\be
{\hat \sigma}_{1b}^2(\epsilon) = - \epsilon + \left[ 1 + {{\hat \alpha} (\alpha
- {\hat \alpha}) \over 1 + {\hat \alpha}^2 } \right]^2 {\hat
\sigma}_1^2,
\ee
and similarly for ${\hat \sigma}_{2b}(\epsilon)$.  The factor ${\cal
J}(\epsilon)$ is a Jacobian factor given by
\be
{\cal J}(\epsilon) = \left( 1 - { \epsilon \over f_1^2 {\hat
\sigma}_1^2} \right)^{-1/2} \left( 1 - { \epsilon \over f_2^2 {\hat
\sigma}_2^2} \right)^{-1/2},
\ee
where $f_1$ and $f_2$ are given in terms of $\epsilon$ by Eqs.\
(\ref{alphadef1}) and (\ref{peak}).

We next insert the result (\ref{ans4}) for $p({\bf
x}|\epsilon)$ into Eqs.\ (\ref{Lambdaformula0}) and
(\ref{Lambdaformula}).  The result is 
\begin{eqnarray}
\Lambda({\bf x}) &=& \int d\epsilon p^{(0)}(\epsilon) { p_\sigma[ {\hat
\sigma}_{1b}(\epsilon), {\hat \sigma}_{2b}(\epsilon) ] \over 
p_\sigma[ {\hat \sigma}_{1b}(0), {\hat \sigma}_{2b}(0) ] }
{{\cal J}(\epsilon) \over {\cal J}(0)} \nonumber \\
\mbox{} && \times 
\exp \left[ - { N \over 2 (1 + {\hat \alpha}^2)} (\alpha^2 - 2 \alpha
{\hat \alpha}) \right].
\label{ans5}
\end{eqnarray}
Finally, integrating over $\epsilon$ gives
\begin{eqnarray}
\Lambda({\bf x}) &=& \sqrt{2 \pi (1 + {\hat \alpha}^2) \over N} 
p^{(0)}({\hat \epsilon}) { p_\sigma[ {\hat
\sigma}_{1b}({\hat \epsilon}), {\hat \sigma}_{2b}({\hat \epsilon}) ] \over 
p_\sigma[ {\hat \sigma}_{1b}(0), {\hat \sigma}_{2b}(0) ] }
{{\cal J}({\hat \epsilon}) \over {\cal J}(0)} \nonumber \\
\mbox{} && \times  \exp \left[  {N {\hat \alpha}^2 \over 2 ( 1 +
{\hat \alpha}^2) } \right] \, \Theta({\hat \alpha}),
\label{ans5a}
\end{eqnarray}
and invoking the arguments of the Secs.\ \ref{ss:sdwnkw} and
\ref{ss:sdwnkwu} above for 
neglecting the prefactors gives 
\be
\Lambda({\bf x}) \simeq \exp \left[  {N {\hat \alpha}^2 \over 2 ( 1 +
{\hat \alpha}^2) } \right] \Theta({\hat \alpha}).
\label{oceans11}
\ee
Thus, in the limit ${\hat \alpha} \ll 1$, $\Lambda({\bf x})$ is
equivalent to the usual cross-correlation statistic $\Theta({\hat
\alpha}) {\hat \alpha}$ defined by Eqs.\ (\ref{hatsigma1def}), 
(\ref{hatsigma2def}), (\ref{hatepsilondef}), and (\ref{alphadef1}).

We end this subsection by recapitulating the various approximations
necessary to obtain the result:

\begin{itemize}

\item The large $N$ approximation $N \gg 1$, necessary for the
validity of the Laplace approximation in integrating over $\sigma_1$,
$\sigma_2$.  

\item The assumption that we are in the regime where the signal is
detectable, $\exp[ N {\hat \alpha}^2/2] \gg 1$.  This is necessary for
the evaluation of the integral over $\epsilon$ in Eq. (\ref{ans5}),
and for neglecting the prefactors in deriving Eq.\ (\ref{oceans11}).

\item The assumption, as before, that the prior probability distributions
$p^{(0)}(\epsilon)$ and $p_\sigma(\sigma_1,\sigma_2)$ are slowly
varying. 

\item The weak signal approximation ${\hat \alpha} \ll 1$, which will
be satisfied unless the stochastic background contribution to the
output of one of the detectors becomes comparable to the noise in that
detector.  As discussed above, for signal strengths at the margin of
detectability, and for several month searches for a stochastic
background with ground based interferometers, 
we have ${\hat \alpha} \sim 10^{-4}$.

\end{itemize}

\subsection{Two coincident, co-aligned detectors, white non-Gaussian
noise}
\label{ss:stochas3}

We next turn to the non-Gaussian noise model of Sec.\ III.A. of paper
I.  The noise in each detector is assumed to be white, with each
sample statistically independent and identically distributed, so that
\be
p_{\bf n}({\bf n}_1,{\bf n}_2 ) = \prod_j
\exp \left[ -f_1(n_{1j}) - f_2(n_{2j})
\right].
\label{stochasnoise2}
\ee
However, it is clear that we cannot assume that the noise
distributions $e^{-f_1}$ and $e^{-f_2}$ in each detector are known in
advance.  Otherwise, as explained in Sec.\ \ref{ss:stochas1}, the
analysis would predict that one can measure the stochastic background
in a single detector by measuring the noise distribution and
subtracting from it the ``known'' noise distribution.

Therefore, in this subsection, we will allow the functions $f_1(x)$
and $f_2(x)$ to be arbitrary except for the normalization conditions
\be
\int e^{-f_1(x)} dx = \int e^{ -f_2(x)} dx = 1.
\label{norm1}
\ee
Formally, there are an infinite 
number of parameters to specify to determine the functions $f_1$ and
$f_2$.  However, in practice these distributions will
be measured as histograms, determined by a finite set of numbers or
parameters.  We identify this finite set of parameters with the noise
parameters $\bftheta_n$ of Eq.\ (\ref{noisePDFexpand}).  
We rewrite Eq.\ (\ref{noisePDFexpand}) as
\be
p_{\bf n}({\bf n}_1, {\bf n}_2) = \int {\cal D}f_1 \int {\cal D}f_2 \,
p_{\bf n}({\bf n}_1, {\bf n}_2 | f_1,f_2) \, p_f[ f_1,f_2].
\label{fnal}
\ee
Here for simplicity we have used a functional or path integral
notation for the integral over the noise parameters (even though the
integral is only over a finite number of parameters).  In Eq.\
(\ref{fnal}), $p_f[f_1,f_2]$ is the prior probability density
functional for the functions $f_1$ and $f_2$, and 
$p_{\bf n}({\bf n}_1, {\bf n}_2 | f_1,f_2)$ is given by the expression
on the right hand side of Eq.\ (\ref{stochasnoise2}).
The probability distribution $p_f[f_1,f_2]$ encodes our assumption
that the noise distributions will have central Gaussian regions, with
unknown variances $\sigma_1$ and $\sigma_2$, and arbitrary tail regions
containing a small fraction $p_{\rm tail}$ of the total probability.

We now insert the distributions (\ref{stochassignal}) and
(\ref{fnal}) into Eqs.\ (\ref{finn}) and
(\ref{noisePDFexpand}) and expand to second order in $\epsilon$.  The
result is 
\begin{eqnarray}
p({\bf x} | \epsilon) &=& \int {\cal D}f_1 \, {\cal D} f_2
\ p_f[f_1,f_2] 
\prod_j e^{-f_1(x_{1j}) - f_2(x_{2j})} \nonumber \\
\mbox{} && \times \prod_j \left[ 1 + \epsilon {\cal A}_j + \epsilon
{\cal C}_j + \epsilon^2 {\cal E}_j + O(\epsilon^3) \right],
\label{ans20}
\end{eqnarray}
where
\be
{\cal A}_j = {1 \over 2} \left[ f_1^\prime(x_{1j})^2 +
f_2^\prime(x_{2j})^2 - f_1^{\prime\prime}(x_{1j}) -
f_2^{\prime\prime}(x_{2j}) \right],
\label{calAdef}
\ee
and
\be
{\cal C}_j = f_1^\prime(x_{1j}) f_2^\prime(x_{2j}).
\ee
The quantity ${\cal E}_j$ is a sum of terms of the form
$f_1^{(n)}(x_{1j})^m$, $f_2^{(n)}(x_{2j})^m$, and 
$f_1^{(n)}(x_{1j})^m \, f_2^{(l)}(x_{2j})^r$ for integers $n,m,l,r$,
whose exact form will not be needed here.

Working to second order in $\epsilon$, we can re-express Eq.\ (\ref{ans20})
as 
\begin{eqnarray}
p({\bf x} | \epsilon) &=& \int {\cal D} f_1 \, \int {\cal D} f_2
\ p_f[f_1,f_2]
  \nonumber \\
\mbox{} && \times 
\exp \left[ - \Xi(\epsilon, f_1, f_2) \right],
\label{ans21}
\end{eqnarray}
where
\begin{eqnarray}
\Xi(\epsilon, f_1, f_2) &=& \Xi_0[f_1,f_2] + \epsilon \Xi_1[f_1,f_2] +
\epsilon^2 \Xi_2[f_1,f_2]
\nonumber\\
\mbox{} &&  + O(\epsilon^3),
\label{Xidef1}
\end{eqnarray}
where
\be
\Xi_0[f_1,f_2]  = - \sum_j
f_1(x_{1j}) + f_2(x_{2j}),
\ee
\be
\Xi_1[f_1,f_2]  = \sum_j ({\cal A}_j + {\cal C}_j),
\label{Xi1def}
\ee
and
\be
\Xi_2[f_1,f_2]  = \sum_j \left[ {\cal E}_j - {1 \over 2} ({\cal A}_j +
{\cal C}_j)^2 \right].
\ee

We now note that we can eliminate the auto-correlation terms ${\cal
A}_j$ from Eq.\ (\ref{Xi1def}) by making a change of variables.
We define the operator ${\cal P}_\epsilon$ that acts on functions via
\be
({\cal P}_\epsilon f)(x) = f(x) - {1 \over 2} \epsilon f^\prime(x)^2 +
{1 \over 2} \epsilon f^{\prime\prime}(x),
\label{calPdef}
\ee
and we define the functions $F_1$ and $F_2$ by
\be
F_1 = {\cal P}_\epsilon f_1, \ \ \ \ \ \ F_2 = {\cal P}_\epsilon f_2.
\label{changevars}
\ee
Using Eqs.\ (\ref{calAdef}),
(\ref{Xidef1})--(\ref{Xi1def}), (\ref{calPdef}) and (\ref{changevars})
the functional $\Xi$ can be rewritten as
\begin{eqnarray}
\Xi(\epsilon, f_1, f_2) &=& \Xi_0[F_1,F_2] + \epsilon {\tilde
\Xi}_1[F_1,F_2] + \epsilon^2 {\tilde \Xi}_2[F_1,F_2]
\nonumber\\
\mbox{} &&  + O(\epsilon^3).
\label{Xians}
\end{eqnarray}
Here the first order piece ${\tilde \Xi}_1$ consists only of the
cross-correlation term,
\be
{\tilde \Xi}_1[F_1,F_2] = \sum_j F_1^\prime(x_{1j})
F_2^\prime(x_{2j});
\label{tildeXi1def}
\ee
the corrections to this cross-correlation expression due to changing
from $f_1$, 
$f_2$ to $F_1$, $F_2$ do not appear at this (linear) order in
$\epsilon$ and instead contribute to the second order term ${\tilde
\Xi}_2[F_1,F_2]$.  The exact form of the functional ${\tilde
\Xi}_2[F_1,F_2]$ will not be needed for our arguments below.

We define the functions ${\hat f}_1(x)$ and ${\hat f}_2(x)$ to be the
functions corresponding to the measured noise distributions at the two
detectors. 
That is, they are step functions defined by the requirement
\be
\int_{-\infty}^x e^{-{\hat f}_1(u) } du = {1 \over N} \sum_{
j {\rm \ with\ } x_{1j} \le x} 1,
\label{hatfdef}
\ee
with a similar equation for ${\hat f}_2$.

Consider now the evaluation of the integral (\ref{ans21}) with $\Xi$ given
by the expression (\ref{Xians}).  Consider first the $\epsilon \to 0$
limit.  In this limit one can show from the normalization conditions
(\ref{norm1}) that $\Xi$ will be minimized at the measured noise
distributions:
\be
F_1 = {\hat f}_1,\ \ \ \ \ \ F_2 = {\hat f}_2.
\label{min1}
\ee
For nonzero $\epsilon$, the leading order correction to the
$\epsilon=0$ result will be given by evaluating the function (\ref{Xians})
at the local minimum (\ref{min1}).  We thus arrive at
\begin{eqnarray}
p({\bf x} | \epsilon) &=& {\cal J}({\bf x}) \, p_f[{\cal P}_\epsilon^{-1}
{\hat f}_1, {\cal P}_\epsilon^{-1} {\hat f}_2] \, 
\exp \left\{ - \Xi_0[{\hat f}_1, {\hat f}_2] \right\}
\nonumber \\
\mbox{} && \exp \left\{  - \epsilon
{\tilde \Xi}_1[{\hat f}_1, {\hat f}_2] - \epsilon^2 {\tilde{\tilde
\Xi}}_2[{\hat f}_1, {\hat f}_2] \right\}.
\label{ans17}
\end{eqnarray}
Here ${\cal J}({\bf x})$ is a width factor whose origin is
approximating the integrals over $f_1$ and $f_2$ as the value of the
integrand at the peak times the ``width'' of the peak \footnote{In
order that the integrand in Eq.\ (\ref{ans21}) be sharply peaked, it is
necessary to make the number of parameters specifying the functions
$f_1$, $f_2$, ${\hat f}_1$, and ${\hat f}_2$ considerably less than the
number $N$ of data points, 
by modifying Eq.\ (\ref{hatfdef}) to incorporate a suitable
coarse-graining of the binning.  This modification does not affect our
argument.}.  This factor is
analogous to the various factors that appear in front of the
exponential in Eq.\ (\ref{ans4}).  It depends weakly on ${\bf x}$ in
comparison to the exponential factors.  
The second order functional ${\tilde {\tilde \Xi}}_2$ in Eq.\
(\ref{ans17}) will differ from the corresponding functional ${\tilde
\Xi}_2$ in Eq.\ (\ref{Xians}) since the location of the peak of the
integrand will receive a correction of order $\epsilon$ away from the
value (\ref{min1}) which will give a correction of order
$O(\epsilon^2)$ to the value of the integral.

We now insert the formula (\ref{ans17}) for $p({\bf x},\epsilon)$ into
Eqs.\ (\ref{Lambdaformula0}) and (\ref{Lambdaformula}).  This gives
\begin{eqnarray}
\Lambda({\bf x}) &=& \int_0^\infty d \epsilon p^{(0)}(\epsilon) {
p_f[{\cal P}_\epsilon^{-1} {\hat f}_1, {\cal P}_\epsilon^{-1} {\hat
f}_1] \over p_f[{\hat f}_1, {\hat f}_1] } \nonumber \\
\mbox{} && \times
\exp \left\{  - \epsilon
{\tilde \Xi}_1[{\hat f}_1, {\hat f}_2] - \epsilon^2 {\tilde{\tilde
\Xi}}_2[{\hat f}_1, {\hat f}_2] \right\}.
\label{ans21a}
\end{eqnarray}
Evaluating the integral over $\epsilon$ gives
\begin{eqnarray}
\Lambda({\bf x}) &=& {\sqrt { \pi \over  {\tilde {\tilde \Xi}}_2 }}
p^{(0)}({\hat \epsilon}) {
p_f[{\cal P}_{\hat \epsilon}^{-1} {\hat f}_1, {\cal P}_{\hat
\epsilon}^{-1} {\hat f}_1] \over p_f[{\hat f}_1, {\hat f}_1] } \nonumber \\
\mbox{} && \times
\exp \left\{  { {\tilde \Xi}_1[{\hat f}_1, {\hat f}_2]^2 \over 4
{\tilde{\tilde \Xi}}_2[{\hat f}_1, {\hat f}_2] }\right\} \,
\Theta({\tilde \Xi}_1),
\label{ans22}
\end{eqnarray}
and as before we can neglect the prefactors to give
\be
\Lambda({\bf x}) \simeq
\exp \left\{  { {\tilde \Xi}_1[{\hat f}_1, {\hat f}_2]^2 \over 4
{\tilde{\tilde \Xi}}_2[{\hat f}_1, {\hat f}_2] }\right\} \,
\Theta({\tilde \Xi}_1). 
\label{ans23}
\ee

Now the statistic ${\tilde \Xi}_1[{\hat f}_1, {\hat f}_2]$ defined
by Eqs.\ (\ref{tildeXi1def}) and (\ref{hatfdef}) coincides with the
locally optimal statistic obtained in paper I [Eq. (3.8) of paper I,
specialized to a white stochastic background], except for the
following modification.  One first measures the noise probability
distributions in each detector separately [cf.\ Eq.\
(\ref{hatfdef})].  Then, one computes the generalized
cross-correlation statistic (\ref{tildeXi1def}) using the those
distributions and the measured data.  

From Eq.\ (\ref{ans23}), we see that $\Lambda({\bf x})$ will be
approximately equivalent to the locally optimal statistic 
${\tilde \Xi}_1[{\hat f}_1, {\hat f}_2]$ if the
statistic ${\tilde {\tilde \Xi}}_2[{\hat f}_1,{\hat f}_2]$ has a weak
dependence on the data ${\bf x}$.  To establish this, consider the
limit $p_{\rm tail} \to 0$, where $p_{\rm tail} \ll 1$ is the total
probability in the noise distribution tails.  
In that limit our assumptions imply that
the noise distributions in the two detectors are Gaussians with
unknown variances $\sigma_1$, $\sigma_2$, and therefore the analysis
of this subsection reduces to the analysis of Sec.\ \ref{ss:stochas2}
above.   Therefore we can read off the $p_{\rm tail} \to 0$ limit of the
statistic ${\tilde {\tilde \Xi}}_2[{\hat f}_1,{\hat f}_2]$ by
comparing Eqs.\ (\ref{ans5}) and (\ref{ans21a}) and identifying the
coefficients of $\epsilon^2$ in the arguments of the exponentials.  
We thus obtain
\be
{\tilde {\tilde \Xi}}_2[{\hat f}_1,{\hat f}_2] = { N \over 2 ( 1 +
{\hat \alpha}^2) } {1 \over {\hat \sigma}_1^2 {\hat \sigma}_2^2 } \,
\left[ 1 + O(p_{\rm tail}) \right].
\label{indirect}
\ee
In the limit ${\hat \alpha} \ll 1$ we can neglect the ${\hat \alpha}$
dependence in Eq.\ (\ref{indirect}).
We then see that the ${\bf x}$-dependent fluctuations in the statistic
are suppressed by either the small parameter $p_{\rm tail}$, as 
as in Secs.\ \ref{ss:color} and \ref{ss:sdcnkw} above, or by the 
parameter $1/\sqrt{N}$ governing the size of the fractional
fluctuations of the statistics ${\hat \sigma}_1$, ${\hat \sigma}_2$.  
Thus, we can neglect the ${\bf x}$ dependence of 
${\tilde {\tilde \Xi}}_2[{\hat f}_1,{\hat f}_2]$ in Eq.\ (\ref{ans23}),
and the approximate equivalence of $\Lambda({\bf x})$ and the locally
optimal statistic ${\tilde \Xi}_1[{\hat f}_1, {\hat f}_2]$ follows.

Finally, we remark that we have not analyzed, in this paper, the most
general situation for stochastic signals of separated, non-aligned
detectors with colored 
noise, which was analyzed in Sec. IV.B of paper I \footnote{
For Gaussian noise, the equivalence of $\Lambda({\bf x})$ and the
standard cross-correlation statistic has been demonstrated for
separated, non-aligned detectors with colored noise in Appendix A of
Ref.\ \cite{flanagan93}, in the limit of weak signals.}. 
However, the results we have obtained make it very plausible that, for
that more 
general situation, the Bayesian statistic $\Lambda({\bf x})$ should
again be equivalent to the generalized cross-correlation statistic
derived in paper I.

We also note that our assumption that the stochastic signal be
Gaussian is necessary for our analysis.  Modifying the signal by
making it be non-Gaussian instead of Gaussian would alter Eqs.\
(\ref{Xidef}), (\ref{Xidefnew}) and (\ref{Xidef1}) at
$O(\epsilon^2)$. Therefore, the derivation here does not generalize 
straightforwardly to non-Gaussian stochastic signals, unlike the
corresponding derivations in paper I.  In Ref. \cite{steve} it is
shown that one can find detection techniques tailored to non-Gaussian
stochastic signals that perform better, for such signals, than the
methods considered here.

\section{CONCLUSION}
\label{s:conclude}

The derivation in this paper, from a different framework, of the
detection strategies obtained in paper I gives us increased confidence
in the utility of those strategies.  In addition, the analysis of this
paper has clarified the regime in which we expect the strategies to
work well.  For deterministic signals, data segments to be analyzed
should be long enough that the signal-to-noise squared per data point
be small.  This requirement is easy to satisfy in practice, as
signal-to-noise thresholds are usually in the range $5 - 10$.  In
addition, the strategies will only be close to optimal in the regime
where signals are strong enough to be detectable; this restriction is
unimportant in practice, as the performance of detection statistics in
the regime where signals are far too weak to be detected is not
important.  Finally, for stochastic signals, the signal must be small
compared to the noise in each individual detector, and the total
probability in the tail part of the noise distributions must be small.

\acknowledgments
This research was supported in part by NSF grants PHY-9728704,
PHY-9722189, PHY-9981795, PHY-0071028, NASA grant NASA-JPL 961298, the
Sloan Foundation, and by the Max Planck Society (Albert Einstein
Institute, Potsdam).  We thank Steve Drasco for helpful conversations
and Tom Loredo for useful comments on the manuscript.


\begin{thebibliography}{}
\bibitem{flanagan93} 
\'E. \'E. Flanagan, Phys. Rev. D \textbf{48}, 2389 (1993).  Note that the
second term on the RHS of Eq. (B6) should read $-10 j_1(\alpha)$ rather than
$-2 j_1(\alpha)$, that the sliding delay function shown in Fig.~2 is
incorrect, and that the right hand side of Eq. (2.8) should be divided by
$\pi$.
\end{thebibliography}
\end{document}